\documentclass[a5paper, 11pt]{article}
\usepackage{graphicx}
\usepackage{fullpage}
\usepackage{fancyhdr}
\usepackage{enumerate}
\usepackage{mathtools}
\usepackage{amsmath}
\usepackage{txfonts}

\usepackage[utf8]{inputenc}
\usepackage[swedish]{babel}
\usepackage{amssymb}
\usepackage[bottom=1.5cm,top=2.5cm,left=2.1cm,right=2.1cm]{geometry}
\usepackage{sectsty}
\usepackage[symbol]{footmisc}
\sectionfont{\large}

\begin{document}
\pagestyle{fancy}
\renewcommand{\headrulewidth}{0pt}
\fancyhf{}
\cfoot{\thepage}

\thispagestyle{empty}
\clearpage
\pagenumbering{arabic} 
\Large
\noindent
\begin{center}
\textbf {Cyclotron Maser Cooling\\ towards Coherent Particle Beams}\footnote[1]{The author wishes to acknowledge the support of Sven Kullander, Fredrik Kullander, Kazutake Kohra, Jun Kondo, Koichi Shimoda, Takeshi Takayama and Bo Vanderberg for his research.}\footnote[2]{E-mail: hikegami@rcnp.osaka-u.ac.jp}
\end{center}
\large
\noindent
\begin{center}
\text{Hidetsugu Ikegami}
\end{center}
\footnotesize
\noindent
\begin{center}
\text{Professor Emeritus, Osaka University}
\text{Formerly Director, Research Center for Nuclear Physics, Osaka University}
\text{Honorary Doctorate, Uppsala University}
\end{center}

\newpage
\large
\noindent
\setcounter{page}{1}
\begin{center}
\text{\textbf{Cyclotron Maser Cooling towards Coherent Particle Beams} }
\end{center}
\small
\noindent
\begin{center}
\text{Hidetsugu Ikegami}
\end{center}
\footnotesize
\noindent
\begin{center}
\text{Professor Emeritus, Osaka University}
\text{Formerly Director, Research Center for Nuclear Physics, Osaka University}
\text{Honorary Doctorate, Uppsala University}
\end{center}
\noindent
\footnotesize
\textbf{\textit{Abstract} } \\  This article presents a new particle beam cooling scheme, namely cyclotron maser cooling (CMC). Relativistic gyrating particles, forced by a solenoidal magnetic field over some length of their trajectory, move in a helical path and undergo emission or absorption of radiations stimulated by a resonance RF field. Theoretical and experimental investigations on electron beams indicate that when the action of the RF field exceeds a critical value the beam jumps abruptly to a coherent radiative system undergoing CMC in which most electrons are accumulated to a discrete energy with the same gyration phase. The mechanism of CMC was proved to be an elementary cooling process that is common to dissipative systems consisting of a driving field and oscillators with a stable energy. It leads to the generation of a coherent beam of particles that provides
means to control miscellaneous particle induced reactions. For example, CMC electrons would generate coherent X-ray and gamma-ray photons through coherent inverse Compton scattering of laser radiation. 

\normalsize
\hyphenchar\font=-1
\section*{1. Introduction}

The invention of beam cooling has opened up new ways of reducing drastically the six dimensional momentum-space phase-space volume of a particle beam in the storage-ring [1-8]. Generally, cooling particle beams is similar to cooling matter in that the thermal movement of particles in a beam, i.e., the random relative motion of constituent particles is reduced as matter does at a low temperature. However, the cooling times achieved in the conventional cooling schemes are extremely long even for a weak intensity of particles stored in a ring typically in the range from one second to one hour for a proton beam of a few hundred MeV. It is noted that most cooling schemes of transverse phase-space so far developed are based on the mechanism of spontaneous transfer of the thermal movement of a particle beam to a low temperature reservoir through heat conduction or radiation, and none has been developed based on the stimulated cooling mechanism. From this viewpoint, the most crucial low temperature reservoir will be the monochromatic radiation field.

According to the principle of cyclotron maser cooling (CMC), relativistic gyrating particles are forced to move in a helical path and interact with an RF field over some length of their trajectory within an action time of the field. In this case, incorporating the relativistic effects and the stimulating field action on gyrations of particles reveals remarkable non-linear features in the particle beam. This dramatically enhances the thermal transfer by stimulated emission or absorption of radiations, removing drawbacks of the present cooling schemes such as the slow cooling speed and the limitations on the energy and the intensity of particle beams to be cooled [9-14]. Of particular interest is that CMC is in principle applicable to any kind of charged particles with any energy and results in an extraordinarily rapid cooling in the transverse energy-time phase-space together with the cooling in the longitudinal energy-time phase-space simultaneously [9].

Experimental indications on CMC in the transverse energy have been obtained with two sets of instruments [10-14]. Some promising evidence of the phase bunching is presented. One piece of evidence clearly shows coherent acceleration of electrons in CMC of energy absorption mode. These results on the dynamical features of CMC electrons are well explained together with those on non-cooling and/or heating electrons by the present CMC theory. The theory has shown that the abrupt appearance of CMC is ascribed to the phase transition of electrons to a coherent state and leads to a coherent charged particle beam in which all free or quasi-free particles are in the same phase and energy. This beam can behave in a completely ordered way, in which the action of any particle is correlated with the action of all the others. The coherent particle beam will open new means to control miscellaneous particle induced reactions through the enhancement of reaction rate, particularly in the cases of coherent electron beam, to generate coherent photons of X and $\gamma$-rays [14].

\hyphenchar\font=-1
\section*{2. Basic Concept of CMC}

Beam particles are often fermions and hence a quantum formulation of CMC for the particles somehow should incorporate Fermi statistics. In the most practical cases of CMC, however, spin dependent effects are negligibly small. This makes it possible to develop the CMC theory that leaves the particle spins out of our consideration.

In conventional particle accelerators and storage rings, the current of the beam particles to be cooled is usually not so large and the beam energy is relativistically high. Thus inter-particle interactions through Coulomb repulsion can be disregarded because of its $\gamma^{-2}$ dependence caused by the relativistic time dilation effect in the interacting particle system, where $\gamma$  denotes the relativistic energy factor $\gamma=(1-\beta^2)^{-1/2}$. 

Consider a field space of length $L$ along the longitudinal (z-) direction and particles of rest mass $m_{0}$, charge $q_{0}$, longitudinal velocity $\textbf{v}_\shortparallel$, and gyration velocity $\textbf{v}_\perp$, within the transverse $x-y$ plane under the Lorenz force, $q_0\textbf{v}_\perp\times\textbf{B}_0$, and the RF field of TE mode, $\textbf{E}=\textbf{E}_0exp(i\omega_{RF}t)$ where $i$ denotes the imaginary number. Throughout this chapter SI units are used. Here the amplitude of oscillation or the Larmor radius of gyration has been assumed to be much smaller than the cross-section of the field space as seen in Fig. 1 (Appendix). In the presence of RF field, particles with random gyration phases undergo emission of thermal radiation and the stimulated absorption of the RF power, resulting in their phase bunching. These features raise the possibility of beam particle cooling when there is a stable energy state in the particle system. Before describing the experimental results, some essential points about the nonlinear CMC mechanism are made below.

In a field of finite length, the momentum conservation, which is generally required for the radiation process, needs not to be satisfied exactly along the longitudinal direction. The allowed deviation in the conservation of wave number (momentum) vectors is of the order of $\pm\pi/L$ resulting in the finite-length-limited linewidth $\Delta\omega_{L}$ of the emission curve of radiation particles [15]:

\begin{equation}
\Delta\omega_{L} = \frac{2\pi}{(1 - \beta_\shortparallel/\beta_{RF}) \tau_t}\: ,
\end{equation}

\begin{equation}
 \tau_t \equiv \frac{L}{\beta_\shortparallel c}\: .
\end{equation}

\noindent Here $\tau_t$,\: $\beta_{RF} c$,\: $\beta_\shortparallel c$\: and $\beta_\perp c$\: denote the transit time of particles, the phase velocity of the RF wave, the longitudinal speed and the gyration speed, respectively. The energy factors are defined as $\gamma=(1-\beta_\shortparallel^2-\beta_\perp^2)^{-1/2}$, $\gamma_\shortparallel=(1-\beta_\shortparallel^2)^{-1/2}$ and $\gamma_\perp=(1-\beta_\perp^2)^{-1/2}$. The linewidth $\Delta\omega_{L}$ is the lowest limit of the bandwidth of the emission curve. There are other mechanisms of broadening (e.g., beam particle collisions, particle energy spread) that increase the emission linewidth beyond this fundamental limit [15]. The considerable part of broadening, how-\\ever, comes from the enhanced damping of gyration energy caused by the collective effect in the phase bunched particles.

The emission and absorption line centers are displaced by a small but significant amount even in the classical regime. The two processes do not cancel with each other exactly due to nonlinear particle gyration with the Lorentz-covariant cyclotron frequency,

\begin{equation}
\frac{\omega_{c}}{\gamma}=\frac{eB_0}{\gamma m_0}\: ,
\end{equation}

\noindent where $\omega_c$ is the cyclotron frequency. In this case, the net stimulated emission line-shape, which is proportional to the difference between emission and absorption line-shape functions, is the derivative of the emission line-shape function, pointing to the nonlinear characteristics of radiation dynamics. As a matter of fact, the rate of energy change of radiating particles is mostly governed by the response function $F(2\Omega \tau_0)$ consisting of a linear term of Lorentzian resonance curve and a nonlinear term of its derivative (see Section 5),

\begin{equation}
F(2\Omega \tau_0) =  \frac{1}{1+(2\Omega \tau_0)^2} +  \frac{4a\Omega \tau_0}{\left[ 1+(2\Omega \tau_0)^2 \right]^2}   \,\, ,   
\end{equation}

\begin{equation}
\Omega \equiv \frac{\omega_c}{\gamma} - \omega_\textrm{RF} \,\, ,   
\end{equation}

\begin{equation}
a \equiv \frac{\omega_c \tau_0}{\gamma}(\beta_\perp \gamma_\shortparallel)^2  \,\, ,   
\end{equation}

\noindent under the assumption of a uniform RF field in the cavity. $\tau_0$ denotes the damping time of gyration, and $\Omega$ is the frequency miss-match between the Lorentz covariant cyclotron frequency $\omega_c / \gamma$ and the RF frequency $\omega_\textrm{RF}$. The factor $a$ (hereafter as correlation factor) is defined in the laboratory frame together with all quantities contrary to the previous papers on CMC [10-14]. In a very weakly relativistic case,  $\beta_\perp \ll 1$, $\gamma_\shortparallel = 1$ and $\gamma = 1 + (\beta_\perp^2 /2)$, Eq.(4) reduces to the well-known Schneider's result [16].

Whenever the correlation condition $a > 1 $ is satisfied, the second nonlinear term in Eq. (4) becomes predominant as seen in Fig. 2 (Appendix). The nonlinear term gives stimulated absorption for $\Omega \tau_0 > 0$, i.e., $\gamma < \omega_c / \omega_\textrm{RF}$, where the particles absorb the RF power to increase their energy $\gamma$ so as to yield the resonance $\gamma = \omega_c / \omega_\textrm{RF}$. The emission is favored to yield the resonance in the case $\Omega \tau_0 < 0$. These features are seen in the $F(2\Omega \tau_0)$ curves with $a > 1$ in Fig. 2. This finding indicates the possibility of CMC because the particles with a higher energy $\gamma > \omega_c / \omega_\textrm{RF}$ undergo stimulated emission while the particles with a lower energy $\gamma < \omega_c / \omega_\textrm{RF}$ undergo stimulated absorption under stimulation of the RF field. A semi-quantitative cal-\\culation on this process leads to extremely rapid phase bunching and energy cooling shorter than the transit time of particles in the RF field. 

Some experimental results which provide inspiring indications of CMC are described in the following sections together with theoretical investigations of the simultaneous processes of phase bunching and energy cooling of gyrating particles.

\hyphenchar\font=-1
\section*{3. First Phase Experimental Test}
\subsection*{3.1 Experimental Arrangement}
\noindent The most crucial problems to be tested experimentally are as follows: does CMC work, can particles be cooled, and can the rapid cooling be realized? CMC is in principle applicable to any kind of charged particles with arbitrary energy. As such, experimental tests using low energy electrons will provide indications of CMC, and hence one may deduce the features of CMC applied to high energy electron and ion beams.

The first phase experimental arrangement for testing CMC is in Fig. 3 (Appendix) [10-13] . An electron beam of about 1 $\mu$ A and 10 keV is generated by a thermionic electron gun with an active domain of about 0.5 mm diameter. The gun is located outside a solenoidal magnetic field. The electrons pass through a dual set of electrostatic horizontal and vertical deflectors move towards an orifice of 40 mm diameter on an iron endplate of 15 mm thickness attached to a cylin-\\drical iron yoke at the entrance of a solenoidal coil of 600 mm length and 200 mm inner diameter. The dual deflector system is arranged so as to adjust the position and direction of the electron beam in the fringing magnetic field. This serves to transform adiabatically any desired fraction of the total kinetic energy into transverse kinetic energy with respect to the solenoid axis. Computer calculations of magnetic field distribution and electron trajectories verified that this way of generating a helical motion worked up to the maximum beam energy but resulted in a very broad spectrum associated with the longitudinal momentum. After their passing through the orifice, the gyrating electrons drift along the magnetic field and pass through an RF cavity with rectangular cross section located at the center of the solenoid. Though these series of experiments were carried out with various cavity modes [11], here described are typical results obtained by using a copper cavity which is 100 mm wide, 50 mm high and 200 mm long and resonant at 2.09 GHz in the $TE_{102}$ mode with a quality factor of Q=8000. The RF power fed to the cavity, adjustable between 0.1 $\mu$ W and 400 mW, is monitored by means of a spectrum analyzer connected to the cavity. Over the cavity space the magnetic field is axially homogeneous with a uniformity of $10^{-3}$. The gyrating electrons are collected on a conductive ZnS screen of 25.3 mm diameter, which is placed just behind an RF gird ring attached to the end window of the cavity. The ZnS screen can be biased to measure the longitudinal drift energy of electrons. Behind the cavity, a video camera equipped with a read out system is set to observe the ring-shaped pattern on the ZnS screen. The entire apparatus is evacuated by a turbo-molecular pump to about $6\times10^{-5}$ Pa without any treatment such as baking out for desorption of gasses.

Electrons pass through the RF cavity of an axial length $L2=0.2 m$ only once, which mostly determines the maximum permissible action time or transit time $\tau_t = L/\beta_\shortparallel c$. The damping time of gyration through the collision of electrons with the residual gas was found to be longer as much as two orders of magnitudes than the transit time of electrons in the cavity and thus neglected. The RF field of the $TE_{102}$ mode in the cavity consists of two travelling waves of $\beta_{RF}=1.4$ propagating along the axis. The interaction of electrons with the RF field is thus characterized by two different value of $\omega_{RF}$  due to the Doppler effect. The resulting two response functions in Eq.(4) have to be taken into account for analyzing the present experimental data based on electrons with extremely low longitudinal drift energy of less than 100 eV or $\beta_\shortparallel=0.02$. This consideration is unnecessary in the cases of high energy particles, since they are resonant with only one of the forward- or backward-travelling waves. In the present experiment it is essential to reduce the drift energy of the electrons to observe the CMC effect in the transverse energy. It is because the low drift energy excludes strictly the possibility of the transfer of the longitudinal energy to the transverse energy being consistent with the assumption of the uniform RF field.

To observe resonance, the magnetic field was swept around $B_0=0.075 T$ which corresponds to the non-relativistic cyclotron resonance $\omega_{c}$. A very strong RF power loss was found at $\omega_{RF}=\omega_{c}$, which is most likely caused by stray electrons emitted from the walls of the apparatus generating an avalanche driven by the RF field, so called multi-factoring effect. Any CMC effect near the resonance at $\omega_{RF}=\omega_{c}$ is therefore difficult to observe. The difficulty of resolving the CMC effect due to the absorption of stray electrons is related to the present small scale experiment in which a low energy electron beam and a small drift velocity result in two overlapping response functions. Through the present experiments, only the resonance condition of coherent absorption mode,

\begin{equation}
 \frac{\omega_{c} }{\gamma_{\perp max}}= (1 - \beta_\shortparallel/\beta_{RF}) \omega_{RF}\: ,
\end{equation}

\noindent was carefully investigated. Here $\gamma_{\perp max}$ denotes the maximum value of $\gamma_{\perp}$. It was found that the absorption mode was a lot more clearly observed than the emission mode due to less interference with the stray electron resonances.

\hyphenchar\font=-1
\subsection*{3.2 Observation of Cooling}

\noindent In spite of the problems outlined in Section 3.1, we have obtained some inspiring indications of cooling of the transverse energy-time phase-space.

In the experiments, the transverse kinetic energy was varied from below 10 keV to beyond 20 keV while the longitudinal drift energy was set at 50-300 eV. Thus it is possible to deduce the transverse kinetic energy distribution precisely by measuring the broadening of the Larmor radii of gyrating electrons depicted on the ZnS screen. The supposition that the centers of gyration are unchanged was hence verified experimentally as explained below.

As an illustration of CMC, the Larmor radii arising from heating and absorptive cooling of electrons are shown in Fig. 4 (Appendix) at a fixed RF power of 100 mW in the cases of initial transverse kinetic enerigies of 8, 10 and 12 keV. Specifically, the cyclotron frequency $\omega_{c}/2\pi=2.16$ GHz ($B_0=0.772 T$) was fixed so that electrons of $\gamma_{\perp max}=1.04$ were resonant with the RF frequency $\omega_{c}/2\pi=2.09$ GHz. The longitudinal drift energy was adjusted to be in the range 50-300eV independent of the initial values of transverse kinetic energy. Because of this large spread of longitudinal drift energy, the electrons were expected to consist of two groups, i.e., $\tau_c > L / \beta_\shortparallel c$ and $\tau_c < L / \beta_\shortparallel c$, depending on their drift velocity $\beta_\shortparallel c$, where $\tau_c$ is the cooling time. The main feature of the pictures with RF on in Fig. 4, where the circular patterns are found, is the formation of a discrete circle enclosing the broad band of circles.

For the electrons of $\tau_c > L / \beta_\shortparallel c$, CMC are not expected from discussions in Section 2, which is verified by the inner broad band of circles as seen in the pictures of RF on in Fig. 4. The broad band is caused by the RF field and expands with the increase of initial transverse kinetic energy. The mean radius is almost equal to the initial Larmor radius with RF off. It is thus a crucial problem whether the formation of the broad band of circles is an effect of the possible oscillation of the centers of gyration or a result from axially symmetric broadening of the Larmor radius due to heating. In order to settle the problem, an experiment was performed. A copper baffle was inserted in perpendicular to the transverse plane outside the cavity exit aperture. The baffle served to take a part of the gyrating electrons away. The electrons were laterally shifted by using the deflector system towards the baffle. If the broad circles were formed by electrons gyrating about different centers, the successive removal of the outer periphery would have resulted in visible effects also at the inner periphery. By taking away so much of the broad circle, it was confirmed that the oscillations of centers caused by the RF field were negligible. The RF field thus accelerated or decelerated the electrons transversely without any marked change of the center of curvature. The feature should be associated with the broad band of circles, where the phases of gyration are uniformly distributed with respect to the RF field.

The sharp outer circle in Fig. 4 corresponds to a transverse kinetic energy of 23 keV and its radius is independent of the initial transverse kinetic energy. This energy coincides with the transverse energy at the resonance $\Omega\tau_0= 0$ for a longitudinal drift energy around 50 eV. It can be seen that three initial sets of electrons, 8, 10 and 12 keV, have reached a final energy of 23 keV. Hence, an energy spread of 4 keV has been compressed up to around 50 eV as far as these electrons are concerned. This assumption is reasonable for electrons undergoing stimulated emission or absorption of radiations, because spontaneous or incoherent effects (e.g., intrabeam scattering effect) can be neglected. With a small longitudinal energy and the deduction of the energy spread $\Delta(\beta_\perp \gamma)^{2} / \Delta(\beta_{\perp0}\gamma_0)^{2}\approx\Delta T_\perp /\Delta T_{\perp0} \approx 0.0125$, the cooling time $\tau_c$ is estimated to be $\tau_c \approx (L / \beta_\shortparallel c)/ln(\Delta T_{\perp0}/\Delta T_\perp)< 0.23L/ \beta_\shortparallel c\approx 11$ ns with the action or the transit time of $L / \beta_\shortparallel c\approx50$ ns.

The simultaneous appearance of broad inner band of circles and a sharp outer circle is due to the large spread of longitudinal drift energy. Note that the broad circle and the sharp circle were separately observed in the second phase experiments by reducing the spread of longitudinal energy as shown in Fig. 7 (Appendix). This indicates that the cooling and heating of electrons are controllable. Nonetheless, the electrons with a broad band of longitudinal energies split clearly into two groups -- one of heating and the other of cooling -- which is very curious.

Once electrons gain their transverse kinetic energy through the stimulated absorption of RF energy, the gain of transverse kinetic energy is further intensified by the decrease of drift velocity $\beta_\shortparallel c$ due to the recoil effect in the coherent absorption of backward travelling RF energy. The amount of the recoil energy $\Delta T_\shortparallel=(\beta_\shortparallel/\beta_{RF})\Delta T_{\perp}$ is of the same order of the longitudinal energy $T_\shortparallel$. For instance, the recoil energy is $\Delta T_\shortparallel=70$ eV for a transverse energy gain of $\Delta T_\perp=5$ keV. For these electrons the effective transit time is estimated to be longer than 50 ns. Then the electrons immediately reach the resonance point $\Omega\tau_0= 0$. On the contrary, for electrons with higher drift energy $T_\shortparallel>100$ eV, their transverse kinetic energy is almost unchanged in average due to their insufficient interaction with the RF field. This situation will be explained in Fig. 9(a) (Appendix). However, this clear separation of heating and cooling groups was revealed in an abrupt appearance of the sharp circle unaccompanied by the heating circle observed in the second phase experiment. The significance of these results will be explained in Section 6.

\hyphenchar\font=-1
\subsection*{3.3 Observation of Phase Bunching}

\noindent The principle of CMC so far discussed is based on the mechanism of bunching of phases of gyrating particles with respect to the applied RF field caused by the relativistic effects. In the experiments, however, the indication of phase bunching was obtained only indirectly through the drastic energy gain associated with the outer circles in Fig. 4. In fact, there has been no way of explaining the energy gain and the sharpness of the outer circles without assuming that the electrons are in resonance at a defined phase with the RF field. The arc or spot pattern was not observed on the ZnS screen. This was again caused by the very broad spectrum of the longitudinal energy. Further, this rounding-off effect on the patterns was enhanced by the recoil effect due to the RF power absorbed coherently by the electrons as will be described in Section 5.

The indication of phase bunching in CMC was seen in another experiment where the sharp outer circle was not necessarily concentric with the broad band of inner circles. As the initial center of gyration was shifted away from the central axis of the RF cavity, the center of the sharp circle shifted towards the central axis of the cavity, where the RF field strength was maximum, and the broad band of the inner circle was concentric about the initially shifted center of gyration. This effect can not be explained unless the electrons forming the sharp circle are assumed to have a certain degree of phase correlation with respect to the RF field -- electrons with phase-matched gyration of absorption are always pulled towards the maximum field. In Fig. 5 (Appendix), the magnetic field is directed normal to the paper and the electrons gyrate in the counter clockwise direction. The axis of the solenoid as well as the central axis of the cavity crosses the ZnS screen slightly to the left of the bottom of the RF off circle. The electric field vector of the RF field is in the vertical direction. When the RF field is on, the broad band of circles is always concentric with the initial RF off circle. The sharp circle is however laterally shifted as seen in Fig. 5. The shift tends to be larger when the central axis of the cavity is further.

\hyphenchar\font=-1
\section*{4. Second Phase Experiments -- Observation of Abruptness of CMC}

The second phase experiments were carried out with an improved Model 2 apparatus made at Uppsala University and is shown in Fig. 6 (Appendix) [14]. An electron beam of 0.3 mm diameter and a current of about 4 $\mu$ A was injected into a solenoid of 1.4 m length through a hole of 50 mm diameter in a 40 mm thick iron end plate attached to the cylindrical yoke. A magnetic deflector was arranged so as to adjust the position and direction of the electron beam at the fringing region of the solenoidal field. This served to adiabatically transform any desired fraction, for example 95\% of the total kinetic energy into the gyration energy in the range of 10-20 keV. After their passing through the fringing field, a helical electron beam with a Larmor radius of about 5 mm ran through a longitudinal energy selector that consists of comb/grid slits aligned along the solenoid axis. The system is covered with a DC-biased copper cylinder for shielding. Another DC-biased cylindrical electrode was installed as a diagnostic tool to measure the longitudinal drift energy and also as an electron shutter to remove slow drift electrons by using negative bias voltages. It was possible to generate an electron beam typically of about 0.4 $\mu$ A with a longitudinal energy of 400 eV with a resolution of about 5\%.

Four silver plated aluminum cavities 100 mm wide, 50 mm high and 200 mm long of the $TE_{102}$ mode were arranged to form an 800 mm long cavity of the $TE_{108}$ mode. The cavity was resonant at 2.1 GHz with a quality factor of Q=12,600. The RF power fed to the cavity, adjustable between 0.1 $\mu$ W and 4 W was monitored by means of a spectrum analyzer connected to the cavity. It is worth noting that the cavity was equipped with two end cylinders, with internal diameters of 30 mm and a length of 60 mm to ensure the suppression of RF power leakage. Over the cavity space the magnetic field was uniform to better than $10^{-3}$ except at the exit point, where a fractional disturbance of a few percent was found. The gyrating electrons were collected on a conductive rugged ZnS phosphor screen of 40 mm diameter to observe their Larmor patterns. The screen could be DC biased to measure the longitudinal drift energy as well as the electron current. The entire apparatus was evacuated by a turbo-molecular pump to  $2\times10^{-4}$ Pa without any treatment such as baking out for desorption of gases.

The electrons passed through the 800-mm-long cavity with a lon-\\gitudinal energy of 400 eV only once, which limited the action time to $\tau_t =67$ ns. Their gyration with an energy of $T_{\perp0} =22$ keV was resonant with the backward propagating RF wave and initially had a uniform phase distribution. Special attention was paid to investigate the dependence of dynamics of gyrations on the RF power in order to confirm the sharp pattern suggesting CMC which was observed in the first phase experiments.

The Larmor pattern of electrons observed under the solenoidal field of $B_0=0.076$ T at the phosphor screen for three typical cases where all the conditions were fixed except for the applied RF power is shown in Fig. 7 (Appendix). As an RF power below the critical value predicted by the CMC condition Eq.(45) was applied, a circular broad pattern appeared as seen in Fig. 7(a). However, the first phase experiments confirmed that the RF power did not cause oscillations of the gyration center. The broadness increased in proportion to the RF power but the average radius remained unchanged. This implies that the energy heating was caused by change in the distribution of the gyration phase as will be explained in Section 5.

As the RF power reached about 2 W, the pattern changed abruptly into a sharp ring of $T_{\perp0}=22$ keV as shown in Fig. 7(b). The sharp ring was not accompanied by any additional ring accumulating all electrons at the same energy when the longitudinal energy resolution was better than 5\%. The appearance of the sharp ring corresponds well to the stable state predicted in Fig. 8 (Appendix). The abrupt change of pattern was accompanied by about 0.2\% decrease of the resonance frequency of the cavity associated with the broadening in the RF band spectrum, i.e., the heating of the RF field. This proportion was close to that of a beam electron acceleration load of $0.4 \mu A \times 10 kV=4$ mW to the cavity loss of 2 W resulting in an almost identical proportional change in the cavity characteristics. This clear abruptness of pattern change became more marked with the increase of longitudinal energy resolution up to 0.5\% which influenced the precision of critical RF field $E_0$ through the resolution of action time $\tau_t = L/\beta_\shortparallel c$. These results will be explained in Section 6 as indications of the phase transition of an incoherent system to a coherent radiative system supporting CMC as the critical point.

When the RF power exceeded the critical value by about 1 dB, the sharp ring indicating CMC turned abruptly into the outer (26 keV) and inner (16 keV) rings as seen in Fig. 7(c). The separation of the rings depended sensitively on the extent of excess power and was independent of the initial gyration energy. The appearance of clear rings and their separation were consistent with the splitting levels predicted by Fig. 8.

Through measurements (a) to (c), the electron current measured at the phosphor screen remained almost unchanged. Through the recoil effect of RF photons, a longitudinal energy shift of 40 eV was caused by the initial gyration energy difference of 2 keV, which disturbed the observation of the phase bunched pattern (but it would be practically insignificant at high longitudinal energies). The phase transition that we observed in the CMC experiments indicates the significance of critical action of the stimulating radiation field.

\hyphenchar\font=-1
\section*{5 Mechanism of CMC}
\section*{5.1 Characteristic Damping Time}

\noindent Consider an RF cavity specified by the resonance frequency $\omega_{RF} $ which is defined as the proper frequency of the empty cavity. When a gyrating charged particle is injected into this cavity, the particle will amplify the thermal radiation in the cavity if its gyration frequency $\omega_c/\gamma $is close to $\omega_{RF}$ [17]. The amplified thermal radiation is finally dissipated as the cavity loss. Suppose that a cavity of volume $V_0$ is tuned in the $TE_{01}$ mode with the quality factor Q. In this case the radiation power emitted by one particle $P_1$, the stored energy in the cavity $W$, and the cavity loss $P_L$ are, respectively

\begin{equation}
P_1=q_{0}r\omega_{RF}E_s\: ,
\end{equation}
\begin{equation}
W= \frac{1}{2}\varepsilon_0 V_0 E_s^2\: ,
\end{equation}
and 
\begin{equation}
P_L= \frac{\omega_{RF} W}{Q}\: ,
\end{equation}
where $\varepsilon_0$ is the dielectric constant, $E_s$ is the field strength in the cavity driven by one particle, and the Larmor radius $r$ is assumed to be much smaller than the cavity cross-section.
At the stationary state $P_L=P_1$, we find
\begin{equation}
E_s= \frac{2q_0rQ}{\varepsilon_0 V_0}\: ,
\end{equation}
that is, 
\begin{equation}
P_1= \frac{2(q_0r)^2\omega_{RF}Q}{\varepsilon_0 V_0}\: .
\end{equation}
As many particles are injected, considerable number $N_B$ of particles will be bunched in their gyration phase because of the collective gyration effect mentioned in Section 2. Thus the thermal radiation is amplified in proportion to $N_B^2$, and their energy $T_{NB}$ proportional to $N_B$.
\begin{equation}
P_{NB}= \frac{2N_B^2(q_0r)^2\omega_{RF}Q}{\varepsilon_0 V_0}\: .
\end{equation}
\begin{equation}
T_{NB}= \frac{N_B}{2}\gamma m_0(r\omega_{RF})^2\: .
\end{equation}
Here the contribution of unbunched particles has been disregarded for simplicity.
The damping time $\tau_0$ of thermal radiation in the cavity is given by
\begin{equation}
\tau_0 \approx\frac{T_{NB}}{P_{NB}}=\frac{m_0\varepsilon_0}{q_0^2}\cdot \frac{V_0 \omega_{RF}}{4Q} \cdot \frac{\gamma}{N_B}\: .
\end{equation}
The damping time $\tau_0$ specifies the characteristics of the system of relativistic gyrating particles in the resonating cavity [17]. Confusing debates on CMC have so far been caused by the lack of consideration on the characteristic damping time $\tau_0$ [18]. The field intensity of the thermal radiation amplified by the collective gyration is obtained from Eqs.(11) and (15) as
\begin{equation}
E_{NB} \equiv N_B E_s \approx \frac{m_0\gamma}{2q_0\tau_0}\cdot r\omega_{RF}= \frac{m_0 c}{2q_0\tau_0}\cdot \beta_{\perp}\gamma \: .
\end{equation}

\hyphenchar\font=-1
\subsection*{5.2 Cooling State of the Dissipative System}
\noindent In order to develop a simple and perspicuous argument on the CMC mechanism, the applied fields $\textbf{E}_0$ and $\textbf{B}_0$ are assumed to be uniform along the z-direction. In these fields, all gyration particles are subject to stimulated emission of radiation and stimulated absorption of RF power resulting in the damping of their gyrating energy, which is specified by the characteristic damping time $\tau_0$ of the system. The frequency of the emitted radiation is close to the resonator frequency $\omega_{RF}$. This inevitably perturbs the RF field $E_0$ as indicated in the second phase experiment described in Section 4.

The damping gyrations are expressed as a frictional force $\textbf{p}_\perp/2\tau_0$ besides the Lorentz force $q_0\textbf{v}_\perp\times\textbf{B}_0$ caused by the solenoidal field, and the driving electric force $q_0\textbf{E}$ of the RF field, where $\textbf{p}_\perp$ denotes the transverse momentum of gyration. The equation of motion is
\begin{equation}
\frac{d\textbf{p}_\perp}{dt}+\frac{1}{2\tau_0}\textbf{p}_\perp+q_0\textbf{v}_\perp\times\textbf{B}_0=q_0\textbf{E}_0e^{-i\omega_{RF}t} \: 
\end{equation}
\noindent in the laboratory frame. It is worth noting that the collective dynamical effect has already been taken into account in Eq.(17); this effect is included in the characteristic damping time $\tau_0$ derived in Section 5.1. Through replacing the variable,
\begin{equation}
\textbf{p}_\perp=m_0\gamma \textbf{v}_\perp = m_0c\beta_\perp\gamma e^{-i(\omega_{RF}t+\phi)} \: ,
\end{equation}
Eq.(17) becomes
\begin{equation}
\frac{d(\beta_\perp\gamma)}{dt}-i(\omega_{RF}+\frac{d\phi}{dt})\beta_\perp\gamma+(\frac{1}{2\tau_0}+\frac{iq_0B_0}{m_0\gamma})\beta_\perp\gamma=\frac{q_0E_0}{m_0c}e^{i\phi} \: ,
\end{equation}
which is equivalent to the coupled equations [17],
\begin{equation}
\frac{d(\beta_\perp\gamma)}{dt}+\frac{1}{2\tau_0}\beta_\perp\gamma=K\cos\phi \: , 
\end{equation}
\begin{equation}
K\equiv\frac{q_0E_0}{m_0c} \: ,
\end{equation}
and
\begin{equation}
\frac{d\phi}{dt}=\Omega-\frac{K}{\beta_\perp\gamma}\sin\phi \: 
\end{equation}
provided
\begin{equation}
\Omega\equiv\frac{q_0B_0}{\gamma m_0}-\omega_{RF} \: .
\end{equation}
Eq.(20) is rewritten in the form,
\begin{equation}
\frac{d[e^{t/\tau_0}(\beta_\perp\gamma)^2]}{dt}=2Ke^{t/\tau_0}\beta_\perp\gamma\cos\phi \: . 
\end{equation}
Its general solution is [19]
\begin{equation}
(\beta_\perp\gamma)^2=2Ke^{-t/\tau_0}\int e^{t/\tau_0}\beta_\perp\gamma\cos\phi\:dt+Ce^{-t/\tau_0} \: . 
\end{equation}
In the right hand side of Eq.(25), the first term is a particular solution that has no arbitrary constant and represents a forced gyration driven by the RF field. The arbitrary constant $C$ of the second term contains the phase $\phi_0$ and the amplitude $\beta_{\perp0}\gamma_0$ of intrinsic gyration, which is determined by the initial condition at $t=0$. As the second term contains a damping factor, it becomes smaller with time. If $t$ is above $\tau_0$, the intrinsic gyration will be damped out and the gyration will be only of the forced gyration. In the limit of $t\to\infty$, Eq.(25) becomes
\begin{equation}
(\beta_\perp\gamma)_{t\to\infty}=\cfrac{2Ke^{t/\tau_0}\cos\phi}{\cfrac{d(e^{t/\tau_0})}{dt}}=2K\tau_0\cos\phi \: . 
\end{equation}
Combining Eqs.(20), (21) and (26), we find 
\begin{equation}
[d(\beta_\perp\gamma)/dt]_{t\to\infty}=0 \: , 
\end{equation}
and hence, at $t\to\infty$
\begin{equation}
\Omega \: \beta_\perp\gamma=K\sin\phi \: . 
\end{equation}
Eqs.(26) and (28) provide respectively the amplitude and the phase of gyration at the cooling state ($t\to\infty$) as
\begin{equation}
\beta_\perp\gamma=\frac{2K\tau_0}{\sqrt{1+(2\Omega\tau_0)^2}} \: , 
\end{equation}
and 
\begin{equation}
\phi= \tan^{-1}(2\Omega\tau_0)\: . 
\end{equation}

In CMC, where the resonance $\Omega=0$ is achieved, Eqs. (29) and (30) are reduced to respective simple forms,
\begin{equation}
(\beta_\perp\gamma)_{cmc}=\frac{2q_0E_0\tau_0}{m_0c} \: , 
\end{equation}
\begin{equation}
\phi_{cmc}=0 \: .
\end{equation}
Eqs.(31) and (32) are equivalent to Eq.(16) and indicate that the forced gyration in CMC is the collective gyration with the RF phase.

The above arguments are applied to the present experiments where $\gamma_\shortparallel\cong1$ and $\gamma\cong\gamma_\perp$. In this case, Eq.(29) predicts a relation between the field action $E_0\tau_0$ required for the cooling and the cooled value $(\beta\gamma)_{t\to\infty}$ of gyration momentum as illustrated in Fig. 8 (Appendix) [17, 20] -- under a given RF field action one or two stable states exist and gyrating particles would be cooled towards these states. The most typical case is $(\beta\gamma)_{t\to\infty}=0$ and $E_0\tau_0=0$  which corresponds to the spontaneous radiation cooling. The conventional cyclotron resonance cooling is also located around the same point in the diagram of Fig. 8. However, in the relativistic cases, whether these stable states appear or not depends sensitively on the finite interaction time $t=\tau_i$, and the difference between the initial and final gyration energies.

\hyphenchar\font=-1
\subsection*{5.3 Cooling Dynamics -- Phase Bunching}
\noindent It is indispensable to solve the coupled equations Eqs.(20)-(22) for investigating the cooling dynamics. Through successive treatments of integration by parts of Eq.(25) and substitution of $d(\beta_\perp\gamma)/dt$ and $d\phi/dt$ with Eqs.(20)-(22) respectively, we have [19]
\begin{equation}
\begin{split}
(\beta_\perp\gamma)^2=-(2K\tau_0)^2F(2\Omega\tau_0)+\frac{2K\tau_0\beta_\perp\gamma}{1+(2\Omega\tau_0)^2}(\cos\phi+2\Omega\tau_0\sin\phi)\\-\frac{4aK\tau_0d(\beta_\perp\gamma)/dt}{1+(2\Omega\tau_0)^2}[\sin\phi-\frac{4\Omega\tau_0}{1+(2\Omega\tau_0)^2}(\cos\phi+2\Omega\tau_0\sin\phi)]\\+Ce^{-t/\tau_0}  \: . 
\end{split}
\end{equation}
Here we have used
\begin{equation}
\frac{d(2\Omega\tau_0)}{dt}=-2a\frac{d(\beta_\perp\gamma)/dt}{\beta_\perp\gamma} \:, 
\end{equation}
with
\begin{equation}
a\equiv\frac{\omega_c\tau_0}{\gamma}(\beta_\perp\gamma)^2 (\frac{\gamma_\shortparallel}{\gamma})^2 \: .
\end{equation}
In the case of the present experiments, the correlation factor $a$ becomes
\begin{equation}
a\cong\frac{2\omega_c\tau_0}{\gamma_\perp}(1-\gamma_\perp^{-1}) \: ,
\end{equation}
which is identical to the correlation factor or the phase bunching factor introduced in previous CMC papers [9-14].

If one takes into account the drastic change of frequency mismatch $\Omega$, Eq.(22) can be solved analytically leading to an asymptotic equation of phase bunching [19]. Using Eqs.(34) and (6), we obtain
\begin{equation}
\frac{d(\Omega\tau_0)}{d(\beta_\perp\gamma)}=\omega_c\tau_0\frac{d(\gamma^{-1})}{d(\beta_\perp\gamma)} =-\frac{a}{\beta_\perp\gamma}\:.
\end{equation}
In usual cases of CMC, we have $a\gg1$ and hence the frequency mismatch $\Omega$ varies rapidly with the phase $\phi$ through the change of $\beta_\perp\gamma$. Therefore Eq.(22) may be represented in the form
\begin{equation}
\tau_0\frac{d\phi}{dt}=\Omega\tau_0-\frac{2K\tau_0}{\beta_\perp\gamma}\sin\phi-\frac{a}{\beta_\perp\gamma}\cdot\frac{d(\beta_\perp\gamma)}{d\phi}\:,
\end{equation}
through Eqs.(20-22), we obtain
\begin{equation}
\tau_0\frac{d\phi}{dt}=\Omega\tau_0-\frac{2K\tau_0}{\beta_\perp\gamma}\sin\phi-a\cfrac{\cfrac{2K\tau_0}{\beta_\perp\gamma}\cos\phi-1}{\Omega\tau_0-\cfrac{K\tau_0}{\beta_\perp\gamma}\sin\phi} \:.
\end{equation}
In the practical cases when $\Omega\tau_0\ll(K\tau_0/\beta_\perp\gamma)\ll a$ and within a short time $t\ll\tau_0$, the first and second terms in Eq.(39) can be disregarded leading to a simple integration to obtain
\begin{equation}
\frac{2K\tau_0}{\beta_\perp\gamma}\cos\phi-1=(\frac{2K\tau_0}{\beta_{\perp0}\gamma_0}\cos\phi_0-1) e^{-at/\tau_0} \:.
\end{equation}
Here the phase $\phi$ bunches with the bunching time,
\begin{equation}
\tau_b=\frac{\tau_0}{a}=\frac{\gamma}{\omega_c}(\beta_\perp\gamma_\shortparallel)^{-2} \: ,
\end{equation}
towards the bunched phase given by Eq.(30). This bunching process proceeds within gyration times of about $(2\pi\beta_\perp^2)^{-1}$ as seen in Eq.(41), causing a collective effect on the stimulated emission and absorption of radiations as argued in Section 5.1.

The first phase experiments show that the phase bunching in CMC is in resonance ($\Omega=0$) at a defined phase ($d\phi/dt=0$) with the RF field. This leads to $\phi=0$ for the collective gyration in CMC recalling Eq.(22). For gyrating particles around the CMC state, the rate of momentum change is obtained using Eqs.(20), (21) and (33) under the condition $\phi\approx0$.
\begin{equation}
\frac{d(\beta_\perp\gamma)}{dt}\approx K-\frac{(\beta_\perp\gamma)^2}{2\beta_\perp\gamma\tau_0}\approx \frac{q_0E_0}{m_0c}F(2\Omega\tau_0) \:, 
\end{equation}
\begin{equation}
F(2\Omega\tau_0)\equiv \frac{1}{1+(2\Omega\tau_0)^2} +\frac{4a\Omega\tau_0}{[1+(2\Omega\tau_0)^2]^2} \:. 
\end{equation}
Eq.(42) supports the arguments in Section 2.

\hyphenchar\font=-1
\subsection*{5.4 Cooling Time}
\noindent Sections 5.2 and 5.3 noted that the RF field breeds the forced gyration of particles over the time $\tau_0$ through stimulated absorption of the RF power. The characteristic damping time $\tau_0$ is thus found to describe the correlation time of particles in the RF cavity. In addition, the intrinsic gyration of initial phase $\phi_0$ and momentum $\beta_{\perp0}\gamma_0$ becomes smaller and finally damps out. The damping time of the intrinsic gyration seems to have significance to the cooling time.

In the case of CMC, the coherent RF field $E_{RF}$ in the cavity driven by an external RF oscillator is superposed on the amplified thermal field $E_{NB}$. This field stimulates absorption of the RF power yielding the coherent forced gyration of particles as seen in the particular solution terms of Eqs.(25) and (33). Nonetheless, the damping factor $exp(-t/\tau_0)$ of intrinsic gyration has to be corrected for the effect of emission of radiations stimulated by the field $E_{RF}$ of the driving RF power [21]. The damping factor of intrinsic gyration should therefore be replaced by the cooling factor $exp(-t/\tau_0)$, provided
\begin{equation}
\tau_c\approx\frac{E_{NB}}{E_{RF}+E_{NB}}\tau_0\approx\frac{E_{NB}}{E_{RF}}\tau_0=\frac{m_0c}{2q_0E_{RF}}\beta_\perp\gamma \:, 
\end{equation}
recalling Eq.(16). Eq.(44) provides the lowest and hence the critical CMC value of the RF field strength $E_{RF}$ for cooling corresponding to Eq.(31) with the replacement of $\tau_0$ by $\tau_c$. The critical field $E_c$ for CMC is
\begin{equation}
E_c\tau_c=\frac{m_0c}{2q_0}(\beta_\perp\gamma)_{cmc} \:. 
\end{equation}

We have dealt with the dynamics of damping and cooling during a long time on the contrary to the cases where the final energy of particles is $\gamma_{t=\tau_t}$ instead of $\gamma_{t\to\infty}$. However, the diagram of field action vs. cooled momentum presented in Fig. 8 is still applicable to the cases of finite-length-limited field action through the replacement of $E_0\tau_0$ and $(\beta_\perp\gamma_\perp)_{t\to\infty}$ by $E_c\tau_c$ and $(\beta_\perp\gamma_\perp)_{t=\tau_t}$, respectively, as seen in Eq.(45).

The essential points of the above argument and those in Section 5.1 on the dissipative system were missing in the open controversy over CMC [18, 22-24].

\hyphenchar\font=-1
\section*{6. CMC as a Phase Transition}
We have looked into the CMC dynamics in the analytical treatments. The dynamics are shown visually by solving the coupled equations Eqs.(20)-(22). Fig. 9 (Appendix) illustrates the cooling dynamics in the gyration velocity phase space diagrams for two typical cases.

Fig. 9(a) shows a phase diagram of gyration in the presence of an RF field strength below the critical value. From the diagram, phases of gyration are uniformly distributed initially. The initial gyration has been assumed to be monochromatic for the sake of clear illustration. The phases are very quickly bunched, but they are incomplete and associated with heating, as described approximately in the scheme of Liouville's theorem. Under the present experimental conditions, the number of gyration times is not necessarily equal to all electrons due to the separation between the RF cavity exit and the phosphor plate and to the longitudinal velocity spread. In this case, the heating illustrated in the diagram will be observed as a broad band of Larmor circles. These features are in good agreement with the experimental result shown in Fig.7(a).

Fig. 9(b) shows a phase diagram of gyrations under an RF field over the critical strength. All conditions are same to those of Fig. 9(a) except for the field strength. Sharp Larmor circles have been observed, as seen in Fig. 7(b) and Fig. 7(c). The cooling time $\tau_c$ is estimated to be 11 ns using Eq.(44) with the value $E_{RF}\geq15$ kV/m of the $TE_{01}$ waveguide mode which is equivalent to the field $E_{RF}\geq21$ kV/m of the $TE_{011}$ cavity mode corresponding to the critical RF power of 2 W. The transit time of electrons is about 67 ns, indicating the agreement between the experimental results and the present arguments. These experimental results have also shown consistency between Eq.(16), Eq.(31) and Eq.(45).

The phase bunching was not directly observed in the experiments due to uncertainty of gyration times during the transit between the exit of RF cavity and the phosphor screen. On the other hand, the consistency of Eq.(44) on the cooling time $\tau_c$ with the experimental results clearly indicates coherent acceleration for the CMC electrons. This result provides evidence for the phase bunching in addition to the experiment described in Section 3.3.

Some observed results remain without satisfactory explanation. The most remarkable one is the abrupt appearance of a sharp ring -- CMC pattern observed in the second phase experiments, which differs clearly from the abruptness in the frequency dependence of nonlinear resonance [14]. We have seen that a system of gyrating particles in the RF cavity is specified by the characteristic damping time $\tau_0$  and the incoherent field $E_{NB}$ generated by the particles. The damping time and the field determine the critical value of field action  $E_{NB}\tau_0$ needed for the phase bunching. This implies that whenever an RF field $E_{RF}$ is driven by an external oscillator, the gyrating particle system undergoes cooling towards a coherent radiative system with a cooling time $\tau_c=E_{NB}\tau_0/E_{RF}$ . Hence if the action time of the field is sufficiently long compared to $\tau_c$, the cooling will be fully developed, resulting in a radiative system of particles in the same phase and stable energy state.

In the present experiments, the action time is determined by the transit time $\tau_t$ of electrons gyrating in the RF cavity and thus the extent of cooling will be dependent upon the factor $exp(-\tau_t/\tau_c)=exp(-E_{RF}\tau_t/E_{NB}\tau_0)$ without any discontinuous change in the state of electrons contradicting to the observation results. The discontinuous behavior of thermodynamic quantities and their derivatives is a strong indication of phase transition of a system in general. In the phase space for free or quasi-free particle systems, one may also expect a similar indication. It is important to investigate the phase space volume of gyrating electrons indicating the sharp CMC pattern. The volume of energy-time phase space is identical to that of momentum-\\space phase space for free or quasi-free particles [25], so the former is estimated in the scheme of CMC theory described in Section 5.

With the following parameters of the second phase experiments, $T_{\perp0}=(\beta_{\perp0}\gamma_{\perp0})_0^2 m_0 c^2/2=10$ keV at $t=0$,  $T_\perp=(\beta_\perp\gamma_\perp)^2 m_0 c^2/2=22$ keV in the final state, and $\tau_c=11$ ns, the response function $F(2\Omega\tau_c)$ is estimated to be negligibly small at $t=0$ and hence $C\cong(\beta_\perp\gamma_\perp)_0^2$. The energy spread at $t=\tau_t$ is
\begin{equation}
\Delta(\beta_\perp\gamma_\perp)_{\tau_t}^2=\Delta(\beta_\perp\gamma_\perp)_0^2\cdot e^{-\tau_t/\tau_c} \:,
\end{equation}
where $\Delta(\beta_\perp\gamma_\perp)_0^2m_0 c^2/2$ is the initial energy spread. 

The thermal energy of gyration is
\begin{equation}
kT=\frac{m_0c^2}{2}[\frac{\Delta(\beta_\perp\gamma_\perp)}{\gamma_\perp}]_{\tau_t}^2=\frac{m_0c^2}{8}[\frac{\beta_\perp\Delta(\beta_\perp\gamma_\perp)^2}{(\beta_\perp\gamma_\perp)^2}]_{\tau_t}^2 \:,
\end{equation}
where $k$ is the Boltzman constant. 

The bunched time $\Delta t'$ in the rest frame of gyration is
\begin{equation}
\Delta t'=[\frac{\beta_\perp^2\Delta(\beta_\perp\gamma_\perp)^2}{\gamma_\perp(\beta_\perp\gamma_\perp)^2}\cos^2\phi]_{\tau_t}\cdot \tau_c\:,
\end{equation}
recalling $\Delta t'=\Delta\phi/\omega_c$ and Eq.(30). 

The phase space volume at the CMC state is
\begin{equation}
\begin{split}
kT\cdot\Delta t'=\frac{m_0c^2\tau_c}{8}[\frac{\Delta(\beta_\perp\gamma_\perp)_0^2}{(\beta_\perp\gamma_\perp)_{\tau_t}^2}]^3 \cdot [\frac{\beta_\perp^4}{\gamma_\perp}]_{\tau_t} \cdot e^{-3\tau_t/\tau_c}\\ \cong 7.9\times10^{-33} [\frac{\Delta(\beta_\perp\gamma_\perp)_0^2}{(\beta_\perp\gamma_\perp)_{\tau_t}^2}]^3 \:.
\end{split}
\end{equation}
For the practical cases $\Delta(\beta_\perp\gamma_\perp)_0^2<0.1\cdot (\beta_\perp\gamma_\perp)_{\tau_t}^2$, the numerical value of Eq.(49) is smaller than the Planck constant -- the uncertainty point of the phase space, where the electrons originate to reveal wave characteristics.

The striking features that are evident in the sharp pattern of CMC electrons and its abrupt appearance have been readily explained as a phase transition to a coherent state of gyrating electrons. In such a coherent particle beam, all free or quasi-free particles are in the same phase and energy state and it becomes the state of entire macroscopic particle beam. The beam can behave in a completely ordered way, in which the action of any particle is correlated with the action of all the others yielding the remarkable collective features on its dynamics. Such a coherent electron beam may provide a means of generating coherent X- and  $\gamma$-ray photons through coherent enhanced inverse Compton scattering of laser radiation [14].

\hyphenchar\font=-1
\section*{7. Conclusion}
The author raised a prime principle of the cooling of charged particle beams in this chapter, thereby developing a new theory of this cooling mechanism by taking a semi-classical dynamical approach. It was found that both the phase bunching and the energy cooling of gyrating particles are caused simultaneously under certain conditions of the stimulating RF field through the relativistic gyration effects. This new theory was examined through experiments using two electron devices. The experimental results were in agreement with this new theory, thereby supporting the new cooling scheme of CMC as the phase transition from an incoherent system to a coherent radiative system. The findings suggest the possibility of generating coherent electron beams that provide a new means of generating coherent X- and $\gamma$-ray photons through the coherent inverse Compton scattering of laser radiation.

\newpage
\section*{Appendix}
\hyphenchar\font=-1
\small

\begin{figure}[h!]
\centering
\includegraphics[scale=0.09]{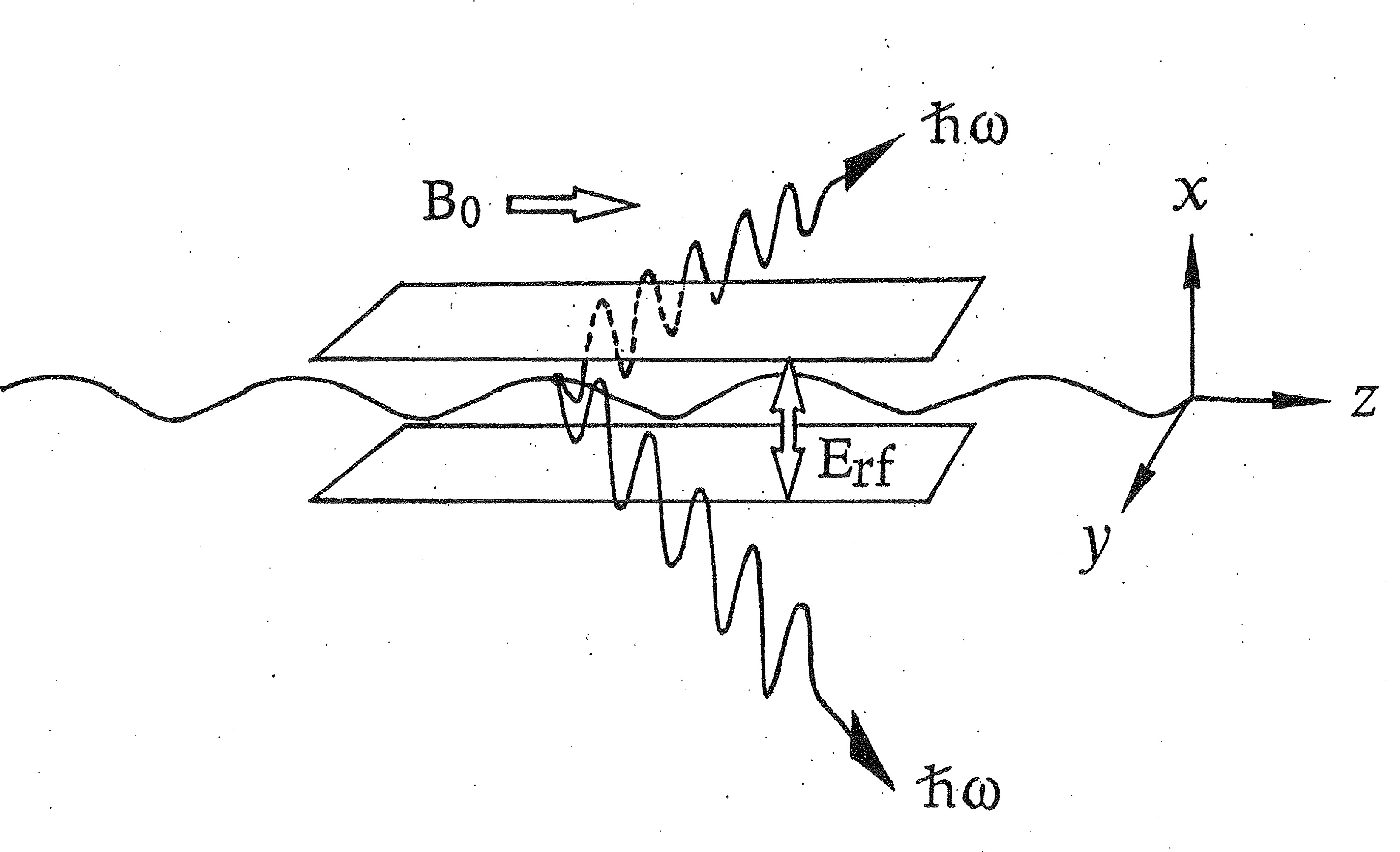}
\end{figure}

\hspace{-0.5 cm}

\noindent 
\small
Fig. 1 \, The conceptual arrangement of the CMC facility. $\textbf{B}_0$: Solenoidal magnetic field; $\textbf{E}_{rf}$: Stimulating RF field of TE-mode; $\hbar\omega$: Stimulated emission photons. The wavy solid line along the z-axis denotes a gyrating particle beam.
\newpage

\begin{figure}[h!]
\centering
\includegraphics[scale=0.1]{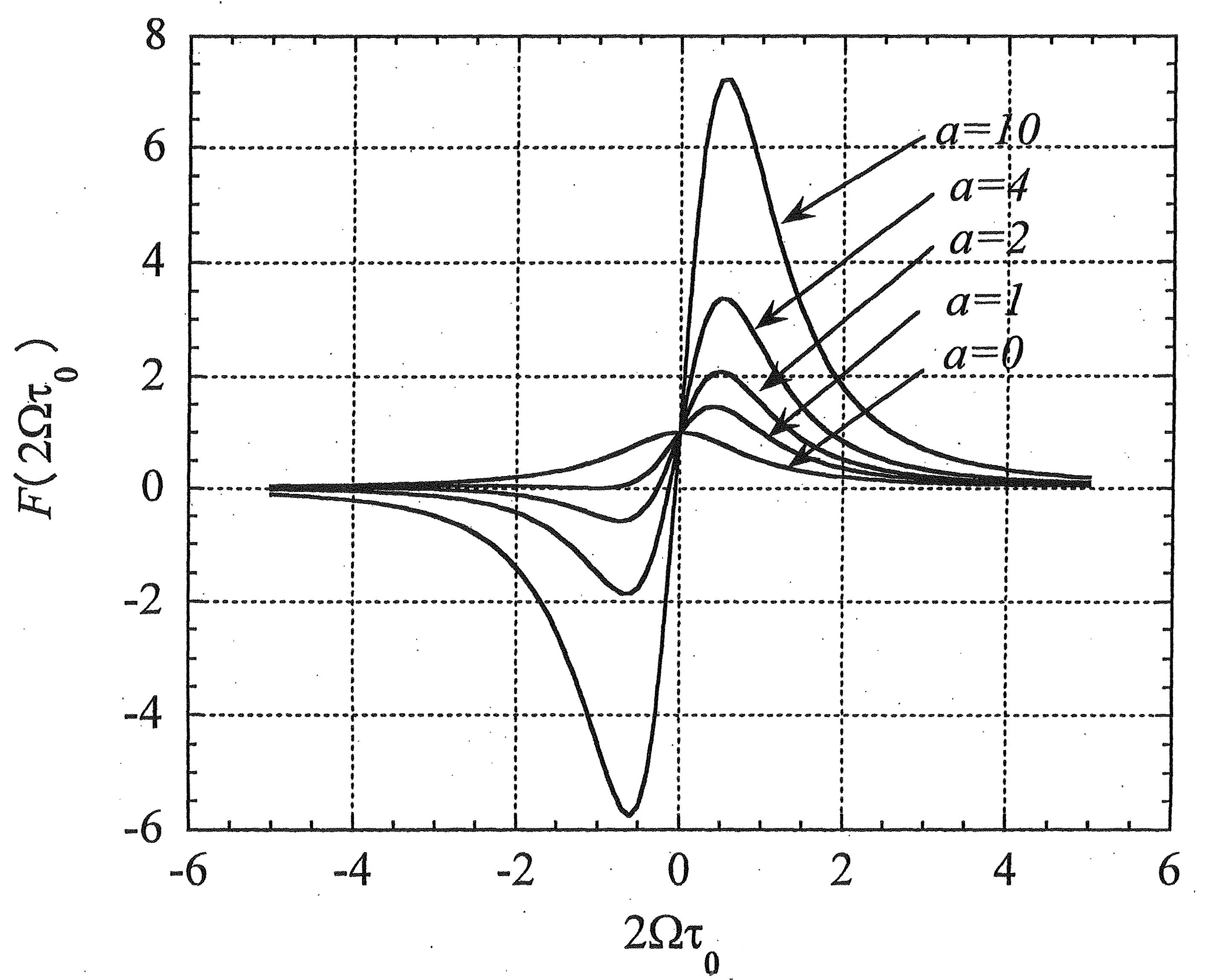}
\end{figure}

\hspace{-0.5 cm}

\noindent 
\small
Fig. 2 \, The response function $F(2\Omega\tau_0)$ gives the normalized energy gain of a gyrating charged particle. For $a\gg1$, the particle with the high energy $\gamma>\omega_c/\omega_{RF}$ emits radiation while the particle with the low energy $\gamma<\omega_c/\omega_{RF}$ absorbs the radiation through the simulation of the RF field resulting in the cooling of gyrating particles.

\newpage

\begin{figure}[h!]
\centering
\includegraphics[scale=0.08]{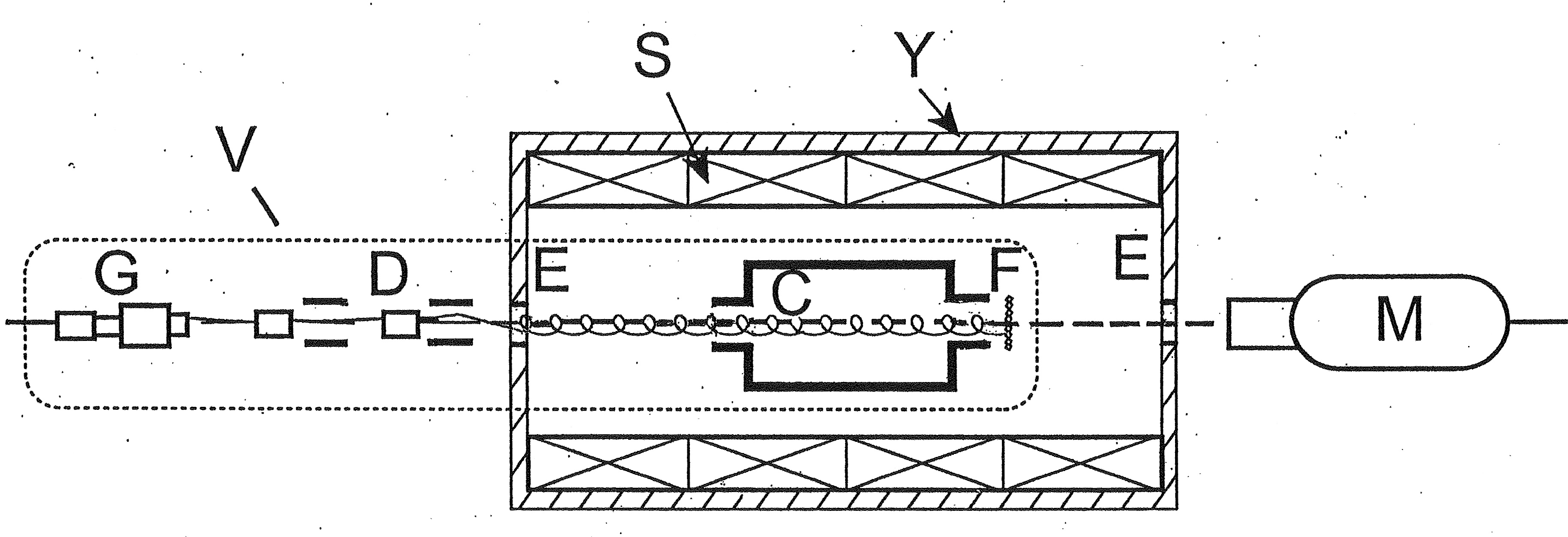}
\end{figure}

\hspace{-0.5 cm}

\noindent 
\small
Fig. 3 \, First phase experimental arrangement (Model 1). G: electron gun; D: dual deflector system; E: end plate of magnet; C: RF cavity; F: conducting ZnS screen; Y: magnet yoke; S: solenoidal coil; V: vacuum enclosure; M: monitoring system.

\newpage

\begin{figure}[h!]
\centering
\includegraphics[scale=0.1]{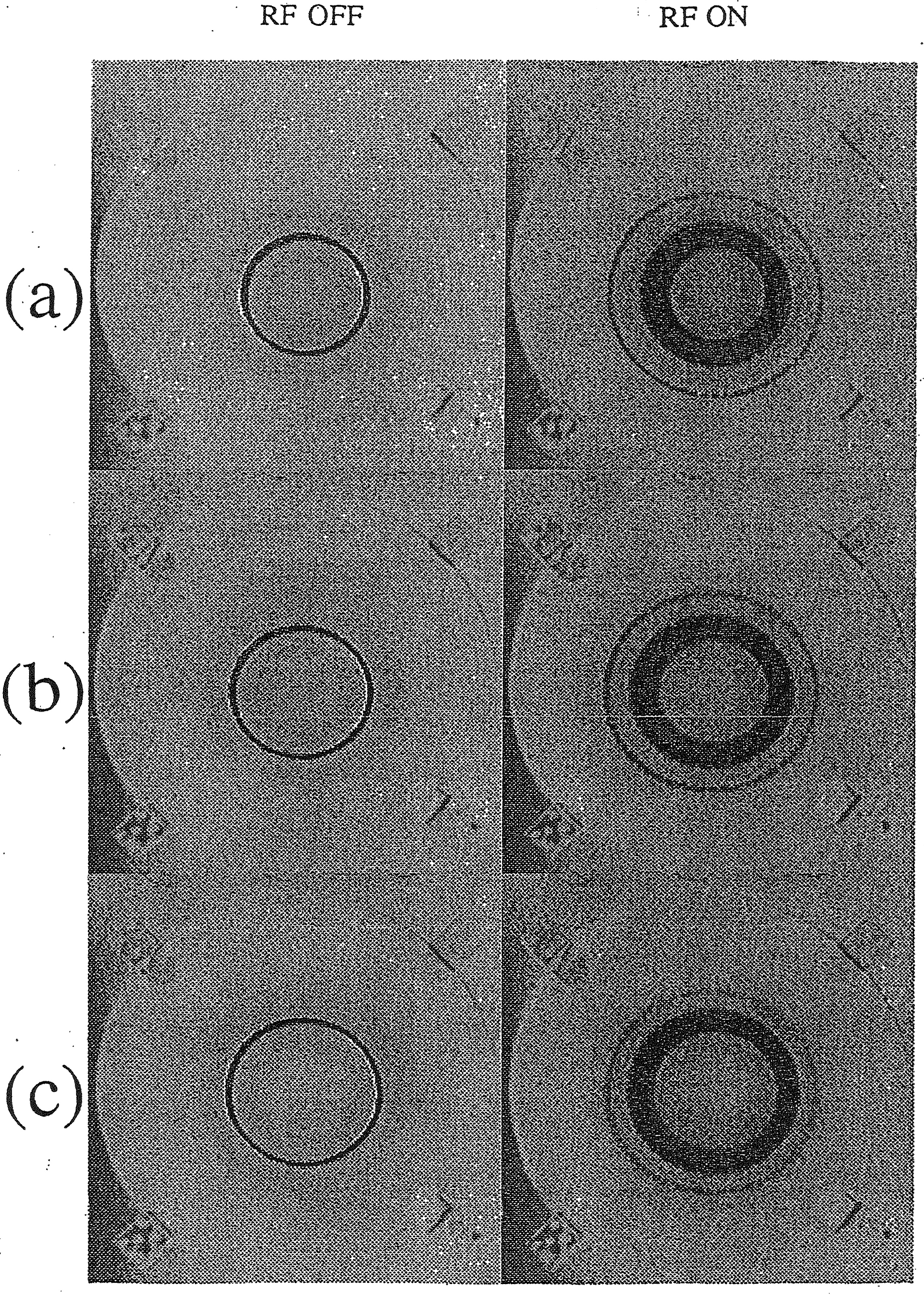}
\end{figure}

\hspace{-0.5 cm}

\noindent 
\small
Fig. 4 \, Comparison of the Larmor radii arising from heating and absorptive cooling of electrons. The patterns depicted on the ZnS screen for RF on and RF off at $\omega_c/2\pi$=2.16 GHz and $\omega_{RF}/2\pi$=2.09 GHz for the initial transverse kinetic energies: (a) 8 keV, (b) 10 keV, (c) 12 keV. The sharp outer circles correspond to an electron energy of 23 keV, which were observed unaccompanied by the broad band of inner circles when the longitudinal drift energy spread was reduced. See also Fig. 7(b). (The pictures are negative.)

\newpage

\begin{figure}[h!]
\centering
\includegraphics[scale=0.125]{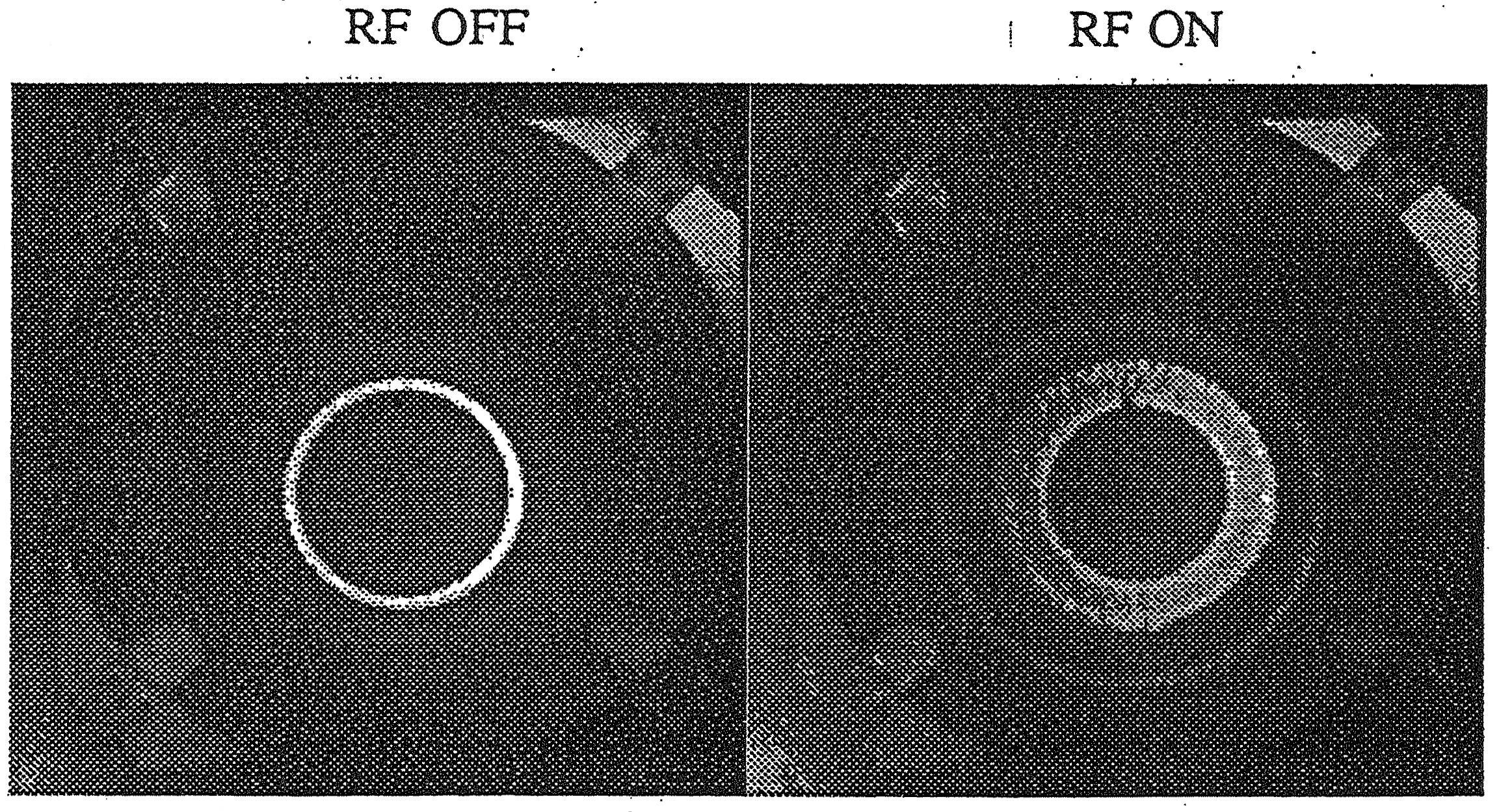}
\end{figure}

\hspace{-0.5 cm}

\noindent 
\small
Fig. 5 \, Indication of phase bunching for the cooled electrons of sharp circle. As the initial center of circle is shifted away from the central axis of the RF cavity, the sharp outer circle with RF on picture shifts back towards the central axis where the strength of the RF field is at a maximum. The broad band of inner circle is symmetric with the initial center of RF off circle. The shift tends to be larger for the further away from the central axis of the RF cavity. The electron beam is prepared with a high transverse to longitudinal kinetic energy ratio. The total energy is 9 keV, cathode current 0.5 $\mu$ A, RF field of 2.091 GHz and 200 mW, cyclotron frequency 2.165 GHz. (The pictures are positive.)

\newpage

\begin{figure}[h!]
\centering
\includegraphics[scale=0.09]{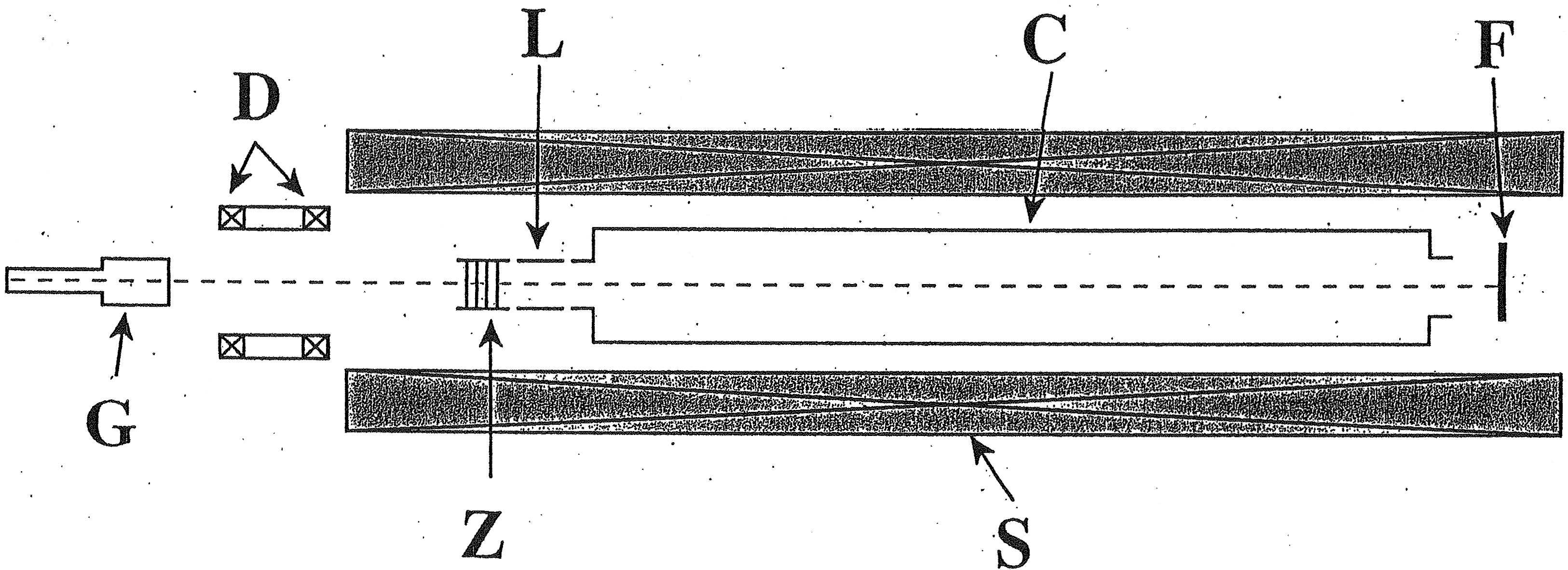}
\end{figure}

\hspace{-0.5 cm}

\noindent 
\small
Fig. 6 \, Second phase experimental arrangement (Model 2). C: RF cavity; D: magnetic deflectors; F: ZnS phosphor screen; G: Electron gun; L: Low energy electron shutter; S: Solenoidal coil; Z: Longitudinal energy selecting comb/grid slits.

\newpage

\begin{figure}[h!]
\centering
\includegraphics[scale=0.1]{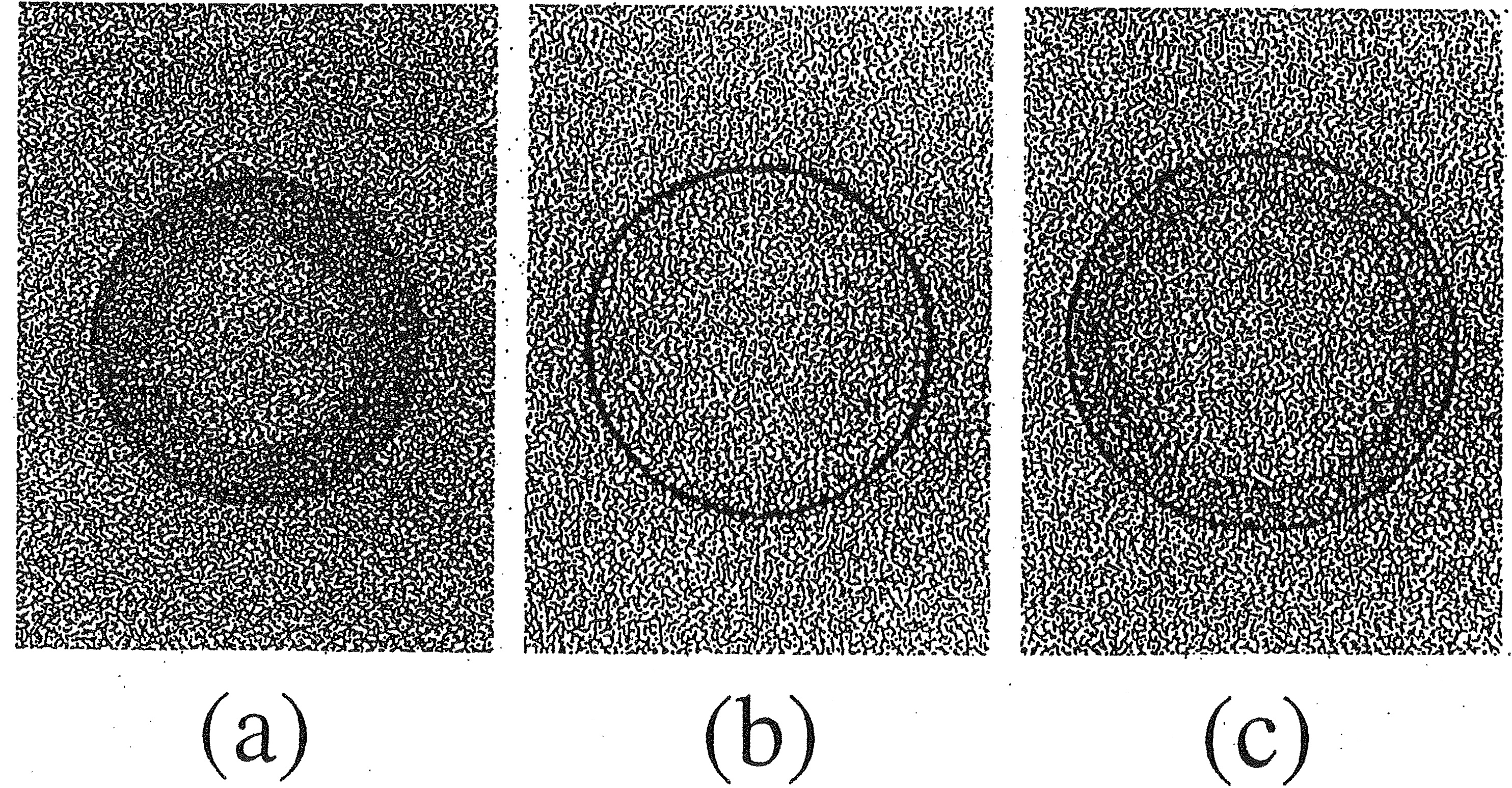}
\end{figure}

\hspace{-0.5 cm}

\noindent 
\small
Fig. 7 \, Comparison of the Larmor patterns of electrons of initial gyration energies $T_\perp$=10.5-12 keV for different stimulating RF power. (a) A pattern observed at 5 dB below the critical power which obeys the Liouville's theorem; (b) The CMC pattern observed at the critical power of about 2 W; (c) A pattern of gyration energy splitting observed at 1 dB over the critical power. All conditions were fixed except for the RF power.

\newpage
\begin{figure}[h!]
\centering
\includegraphics[scale=0.112]{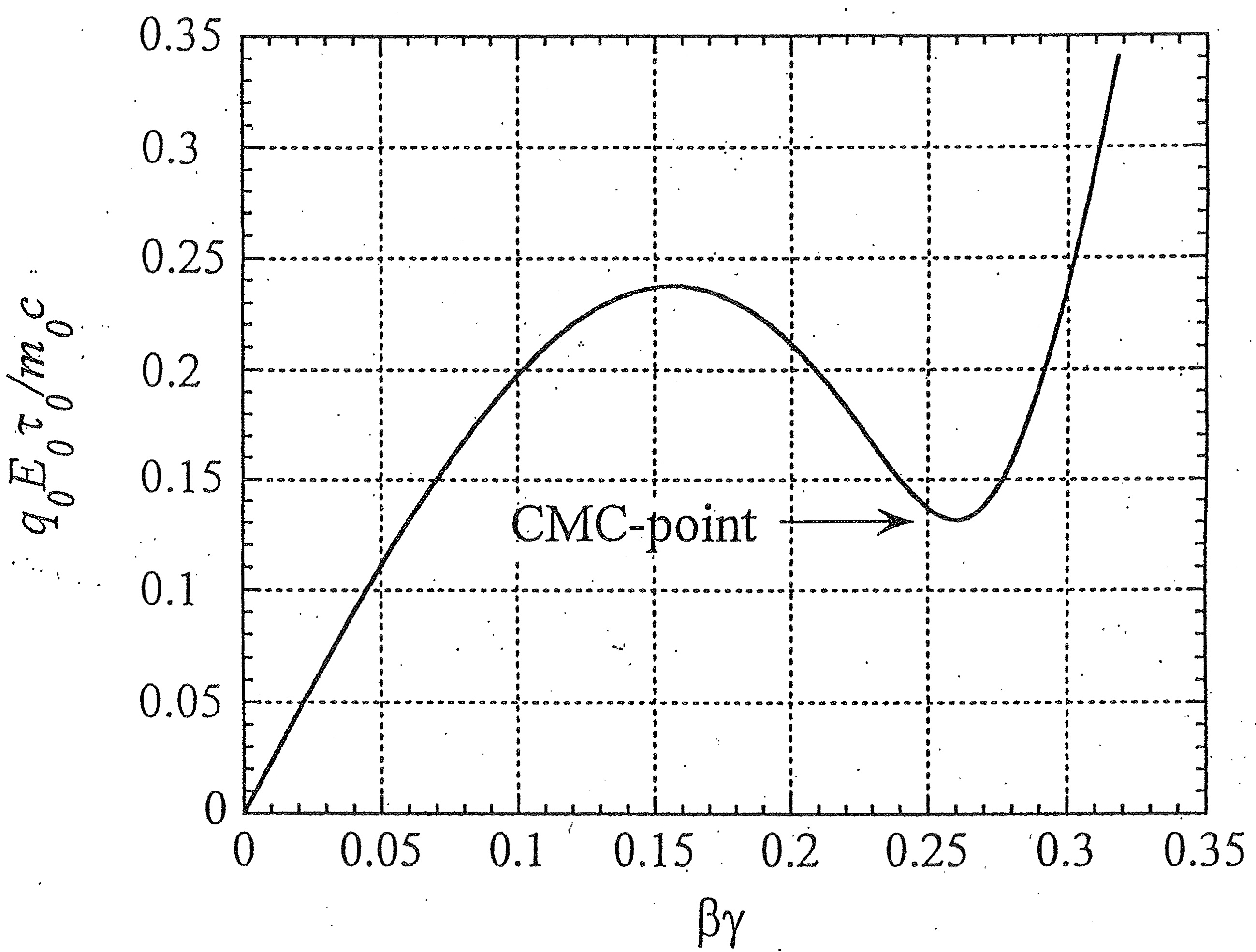}
\end{figure}

\hspace{-0.5 cm}

\noindent 
\small
Fig. 8 \, Field action $E_0\tau_0$ vs. cooled momentum $(\beta_\perp\gamma_\perp)_{t\to\infty}$. Under an RF field action below the critical value, there is no stable state of relativistic gyrating particles. As the field action reached the critical CMC value, a cooled state appears at a relativistic energy. At a field action exceeded the critical value, the stable state splits into two states.

\newpage

\begin{figure}[h!]
\centering
\includegraphics[scale=0.075]{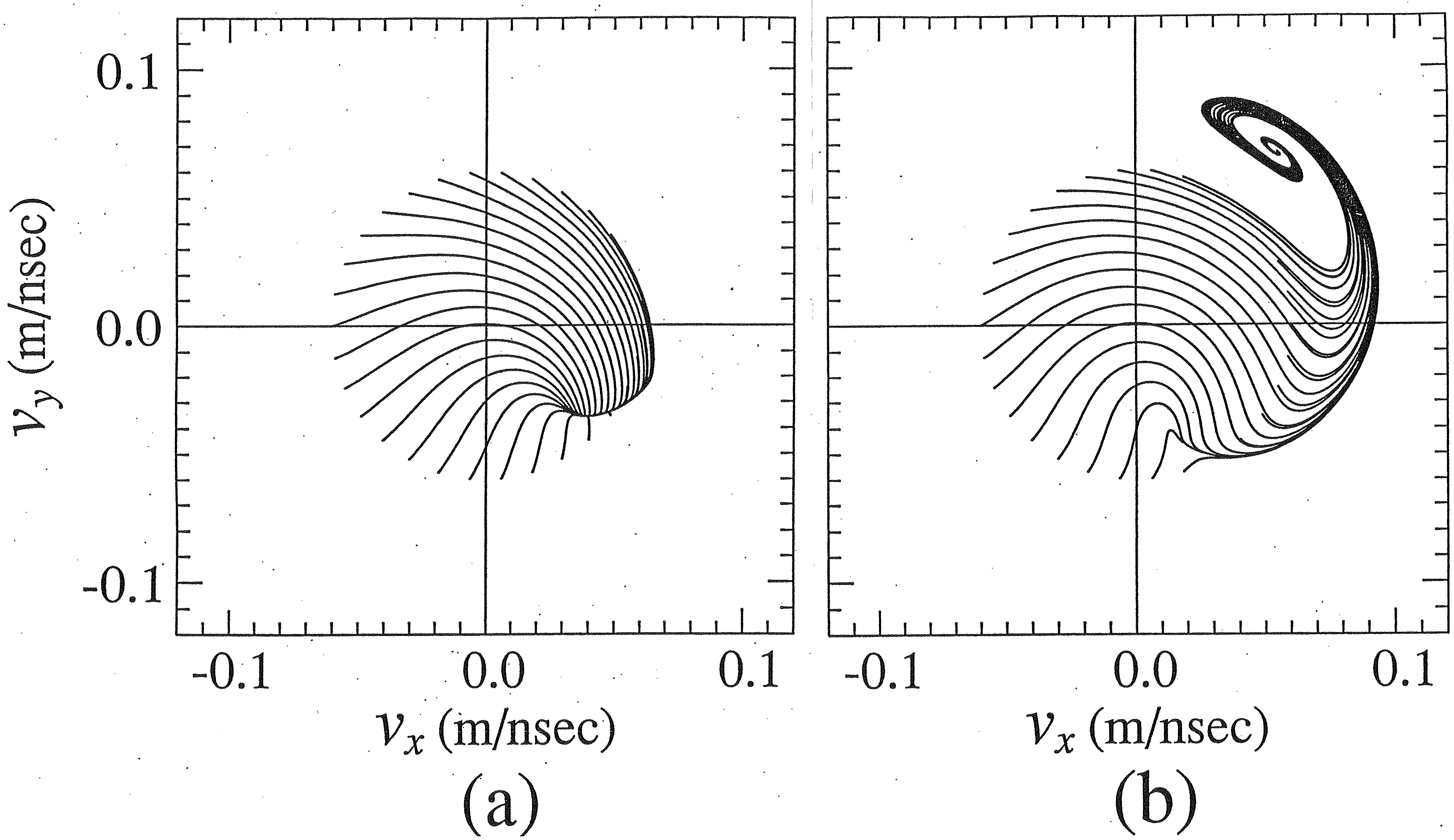}
\end{figure}

\hspace{-0.5 cm}

\noindent 
\small
Fig. 9 \, Phase diagram in a momentum space of particle gyration. (a) Under a stimulating RF field below the critical value prescribed by the CMC condition, all particles with uniformly distributed gyration phase are bunched in phase but heated in the gyration energy as indicated by an arc envelope. This will be observed as a Larmor pattern indicated by a broad band through averaging over the gyration period as shown in Fig. 7(a); (b) Under the RF field satisfying the CMC condition, all of the particles are accumulated in a cooling state of one discrete energy and the same gyration phase. This will be observed as a sharp Larmor pattern as shown in Fig. 7(b).

\newpage
\small
\section*{References}
\begin{enumerate}

\item G.I. Budker et al, Part. Acc. \textbf{7} (1976) 197.

\item G.I. Budker and A.N. Skrinsky, Usp. Fiz. Nauk. \textbf{124} (1978) 561. [Sov. Phys. Usp. \textbf{21} (1978) 277].

\item H. Poth, Phys. Rep. \textbf{196} (1990) 135.

\item A.N. Skrinsky and V.V. Parkhomchuk, Fiz. Elem. Chastits At. Yadra \textbf{12}  (1981) 557. [Sov. J. Part. Nucl. \textbf{12}  (1981) 223], and references cited therein.

\item D. Moehl, G. Petrucci, L. Thorndahl and S. van der Meer, Phys. Rep. \textbf{58} (1980) 73.

\item M. Sands, Proc. Int. School “Enrico Fermi”, 46th Course, Varenna, Italy, ed. B. Touschek, Academic Press, New York, 1969.

\item E. Bonderup, CERN 90-04, ed. S. Turner, (Proc. CERN, Third Advanced Accelerator Physics Course, Uppsala, Sweden, 18-29 September 1989).

\item K. Kilian, Proc. Workshop on Electron Cooling and New Cooling \\Technique, (p.239), eds. R. Calabrese  \& L. Tecchio, World Scientific Press, Singapore, 1991.

\item H. Ikegami, Phys. Rev. Lett. \textbf{64} (1990) 1737; \textbf{64} (1990) 2593.

\item H. Ikegami, Physica Scripta \textbf{48} (1993) 32; (Proc. Second International Conference on Particle Production near Threshold, (Opening Talk), ed. C. Ekstroem.

\item F. Kullander, Cyclotron Maser Cooling, PhD Thesis, Osaka University, 1993.

\item H. Ikegami, Proc. Workshop on Beam Cooling and Related Topics, (p.81), CERN 94-03, ed. J. Bosser, (CERN, Geneva, 1994).

\item H. Ikegami, Photon Radiation \textbf{8} (1995) 253.

\item H. Ikegami, J. Phys. Soc. Jpn. \textbf{66} (1997) 2214.

\item A. Friedman, A. Gover, G. Kurizki, S. Fluscin and A. Yariv, Rev. Mod. Phys. \textbf{60} (1988) 471.

\item J. Schneider, Phys. Rev. Lett. \textbf{2} (1959) 504.

\item K. Shimoda, The first and second CMC Seminars, org. T. Takayama (Unpublished, Tokyo, 1998).

\item D. Moehl and A.M. Sessler, Proc. Workshop on Beam Cooling and Related Topics, (p. 429), CERN 94-03, ed. J. Bosser (CERN, Geneva, 1994), and references cited therein.

\item H. Ikegami, The third and fourth CMC Seminars, org. T. Takayama (Unpublished, Tokyo, 1998, 1999).

\item T. Takayama and J. Kondo, The second and third CMC seminars, org. T. Takayama (Unpublished, Tokyo 1998).

\item H. Ikegami, CMC-Note (Unpublished 1999).

\item J.L. Hirschfield and G.S. Park, Phys. Rev. Lett. \textbf{66} (1991) 2312.

\item D.J. Larson, Phys. Rev. Lett. \textbf{68} (1992) 133.

\item J.L. Hirshfield and G.S. Park, Phys. Rev. Lett. \textbf{68} (1992)134.

\item R. Eisberg and R. Resnick: Quantum Mechanics of Atoms, Mole-\\cules, Solids, Nuclei and Particles, 2nd ed. (Wiley, NewYork, 1974).

\end{enumerate}
\newpage
\large
\begin{center}
\hspace{-0.5cm}
\textsc{About The Author}
\end{center}
\vspace{0.5cm}
\normalsize
\hyphenchar\font=-1
\noindent
Professor Hidetsugu Ikegami was the former Director (1985-1993) of the Research Center for Nuclear Physics (RCNP), Osaka University. He designed and built the World High Resolution Spectrographs \textit{RAIDEN} and \textit{GRAND RAIDEN} equipped with the complex multipole fields tuning and spectrograph-beam line matching systems as well as setting up the RCNP Ring Cyctron Facilities. He received an honorary doctorate from Uppsala University (1990) with his achievements in high precision nuclear spectroscopy. He was made an honorary citizen of Tennessee (1965) for his contribution to Atomic Data and Nuclear Data Tables Reform at Oak Ridge National Laboratory. He became a figure of academic renown by finding the first evidence for “Doorway States” [\textit{Phys. Rev. Lett.} 13 (1964) 26], otherwise known as “Feshbach Resonance”.\end{document}